\documentclass[
    ,final            
  ]
  {aipproc}

\layoutstyle{6x9}

\begin{document}

\title{Precision measurements of positronium decay rate and energy level}

\classification{PACS numbers}

\keywords  {Positronium, QED, decay rate, HFS}

\author{S.Asai, Y.Kataoka, T.Kobayashi, T.Namba, T.Suehara, G.Akimoto, A.Ishida, M.M.Hashimoto}{
  address={ICEPP, Faculty of Science, University of Tokyo, 7-3-1 Hongo, Bunkyo-ku, Tokyo 113-0033}
}

\author{H.Saito}{
  address={Graduate school of Arts and Sciences, University of Tokyo, 3-8-1 Komaba, Meguro-ku, Tokyo, 153-8902}
}

\author{T.Idehara}{
  address={FIR Center, University of Fukui, 3-9-1 Bunkyo, Fukui, 910-8507}
}

\author{M.Yoshida}{
  address={High Energy Accelerator Research Organization (KEK) 1-1 Oho, Tsukuba,Ibaraki, 305-0801}
}

\begin{abstract}

Positronium is an ideal system for the research 
of the bound state QED.
New precise measurement of orthopositronium 
decay rate has been performed with 
an accuracy of 150 ppm,
and the result combined with the last three is
7.0401$\pm$0.0007$\mu s^{-1}$.
It is the first result to validate the 2nd 
order correction.
The Hyper Fine Splitting of positronium 
is sensitive to the higher order corrections 
of the QED prediction and also to 
the new physics beyond Standard Model
via the quantum oscillation into virtual photon.
The discrepancy of 3.5$\sigma$ is found recently between 
the measured values and the QED prediction ($O(\alpha^3)$).
It might be due to the contribution of the new physics or
the systematic problems in the previous measurements:
(non-thermalized Ps and non-uniformity of the magnetic field).
We propose new methods to measure HFS precisely without 
the these uncertainties.

\end{abstract}

\maketitle


\section{Introduction}

Positronium (Ps), the bound state of 
an electron and a positron,
is a purely leptonic system
and the triplet ($1^{3}S_{1}$) state of Ps, orthopositronium(o-Ps),
decays slowly into three photons.
It is good sample to measure decay rate directly and precisely.
The o-Ps has an energy higher than 
the single state ($1^{1}S_{0}$) of Ps, parapositronium(p-Ps).
The difference of the energy level is 
called as Hyper Fine Splitting (HFS) and
is significantly larger(about 203~GHz) 
than the hydrogen-atom (1.4~GHz).
Since o-Ps has the same quantum number of photon,
the quantum oscillation o-Ps $\rightarrow \gamma^{*} \rightarrow$ o-Ps
contributes the HFS (87~GHz), and the oscillation has 
the good sensitivity to the new physics beyond Standard Model.
 
Precise measurements of the decay rate of o-Ps and 
the HFS give us direct information about quantum electrodynamics(QED)
in bound state.
The higher order calculations in the bound state 
have be developed recently, and 
the we have a good chance to compare the experimental 
results with the QED predictions directly.

There are two topics in this note.
We report the first test of
the 2nd order calculation of the orthopositronium 
decay rate in the first part. 
We perform the accurate measurement (error $<$150ppm)
of the o-Ps decay rate, and the 2nd order calculation is directly 
validated for the first time. 
In the second part, we propose the new methods 
to measure the HFS.
There is discrepancy(3.5$\sigma$) between the prediction($O(\alpha^3)$)
and the precise measured values. 
We point out the possible two systematic errors in the previous 
measurements and propose the new methods without these systematic 
uncertainties.

\section{orthopositronium decay rate}

\subsection{History and current status}

Three precise measurements\cite{GAS87,GAS89,CAV90} 
of the o-PS decay rate were performed,
in which reported decay rate values much larger, i.e., 5.2 -- 9.1 experimental 
standard deviations, than a QED prediction\cite{ADKINS-4}
($7.039934(10)~\mu s^{-1}$) corrected up to $O(\alpha^2)$\@.
This discrepancy has been referred as `orthopositronium lifetime puzzle', 
and was long-standing problem.

As some fraction of o-Ps inevitably results in `pick-off' annihilation 
due to collisions with atomic electrons of the target material, 
the observed o-Ps decay rate 
$\lambda_{obs}$ is a sum of the intrinsic o-Ps 
decay rate $\lambda_{{\rm o}\mbox{-}{\rm Ps}}$ and the pick-off 
annihilation rate into $2\gamma$'s, 
$\lambda_{pick}$, i.e.,

\begin{equation}
\lambda_{obs}=\lambda_{3\gamma}+\lambda_{pick}.
\end{equation} 

In the old measurements\cite{GAS87,GAS89,CAV90}, 
$\lambda_{obs}$'s were measured by varying the 
densities of the target materials, size of the cavities and
also the entrance aperture of the cavities.
The extrapolation to zero density or aperture was expected to 
yield the decay rate in a vacuum, $\lambda_{3\gamma}$, under 
the assumption of the constant o-Ps velocity. 
We pointed out\cite{PEKIN,ASAI95} that this assumption was the 
common/serious systematic uncertainties in these results,
since it takes much time that Ps is well thermalized.

We have proposed the following entirely new method\cite{ASAI95}, 
which is free from above-mentioned systematic error.
The energy distribution of photons emitted from the 3-body decay is 
continuous below the steep edge at 511~keV, 
whereas the pick-off annihilation is 2-body which produces 
a 511~keV monochromatic peak. 
Energy and timing information are simultaneously
measured with high-energy resolution germanium detectors such 
that $\lambda_{pick}(t)/\lambda_{3\gamma}$ can be determined
from the energy spectrum of the emitted photon. 
Once a precise thermalization function is obtained, $\lambda_{pick}(t)$ will 
contain all information about the process. 
The population of o-Ps at time $t$, 
$N(t)$ can be expressed as \begin{equation}
N(t)=N_0' \exp\left(-
\lambda_{3\gamma}\int^{t}_0\left(1+\frac{\lambda_{pick}(t')}
{\lambda_{3\gamma}}\right)dt'\right).
\end{equation}
Providing the ratio is determined as a function of time,
the intrinsic decay rate of o-Ps, $\lambda_{3\gamma}$, can be directly obtained
by fitting the observed time spectrum.

We obtained decay rate of $7.0398(29)~\mu$\cite{ASAI95}, 
are consistent with the first order QED calculation,
and quite differ from the previous results as shown in Fig.~1\@.
This result are confirmed by two different and more accurate~(200ppm)
measurements\cite{JIN,MPOL} and the 'lifetime puzzle' has been solved.
Now the interesting is focused 
on the validation of the higher order correction itself.

Non relativistic QED approximation has been developed recently, 
and it is useful to calculate higher oder correction of the bound state.
The second order correction\cite{ADKINS-4}, whose contribution is about 160ppm, 
has been performed in 2002,  and 
the more accurate measurement  is necessary to examine  the 2nd order correction.

\begin{figure}
  \includegraphics[height=.3\textheight]{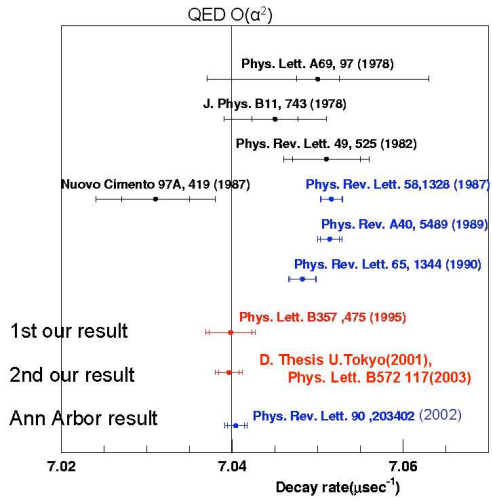}
  \caption{Historical plots of the measured o-Ps decay rate
}
\end{figure}

\subsection{Experimental setup}

Figure~2 shows a diagram of the experimental setup\cite{kataD};
A ${}^{68}{\rm Ge-Ga}$ positron source (dia., $10~mm$) with the 
strength of $0.3\mu Ci$, being sandwiched between two sheets of plastic 
scintillators(NE102 thickness=200$\mu$m) 
and held by a cone made of aluminized mylar. 
The cone was situated at the center of a cylindrical vacuum container
made of $1~mm$-thick plastic scintillators and glass, 
being filled with ${\rm SiO_2}$ aerogel~(RUN-I) or powder~(RUN-II),
and evacuated down to $1\times10^{-2}$ Torr. 
Density of the aerogel and powder is 0.03g/cc both,
and the surface of the primary grain are replaced into hydrophobic
in oder to remove the electric dipole of the OH-. 
The sizes of the primary grain are 10 and 16 nm for the aerogel and powder, respectively.
Thus  the mean free path of the collision between positronium
and ${\rm SiO_2}$ are different,
and it makes difference in the thermalization process and the pickoff probability. 

\begin{figure}
  \includegraphics[height=.4\textheight]{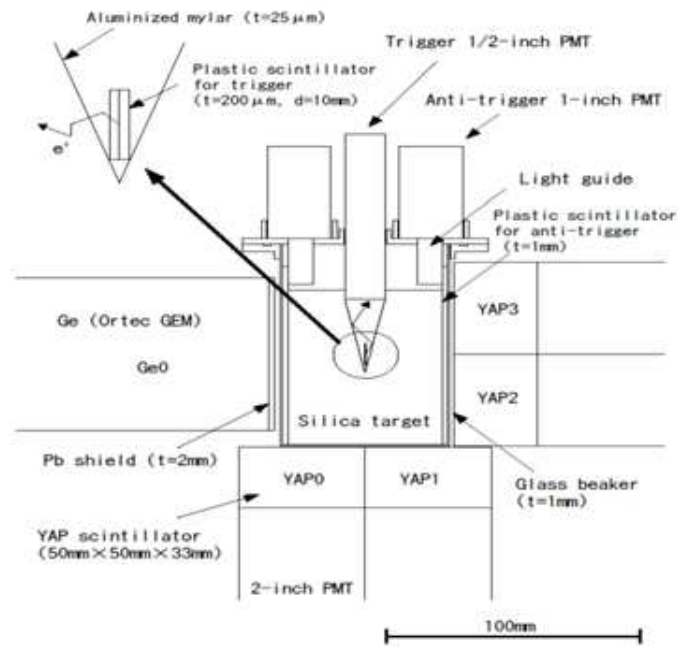}
  \caption{Schematic diagram of apparatus.\cite{kataD}
}
\end{figure}

The emitted positrons pass through thin scintillator,
transmitting a light pulse to a 
photomultiplier(Trigger PMT) 
and forming Ps when stopped in the silica powder or aerogel.
Since the kinetic energy of the positron 
is high ($E_{end}$=1.9MeV: ${}^{68}{\rm Ge-Ga}$)
about 40\% of the positron pass through the silica and annihilate at the 
vacuum container.
If the positron deposits energy(>100~keV) 
in the $1~mm$-thick plastic scintillators, inside wall of 
the vacuum container, the produced trigger signal
is vetoed to remove these annihilate events in the wall.

Three high-purity coaxial germanium detectors (Ortec GEM38195) 
precisely measured the thermalization process.
Energy resolutions were measured using several line $\gamma$ sources, 
with typical resultant values of 0.5~keV at 514~keV. 
Four YAP($\rm {YAlO_{3}}$: Ce doped: size is  50$\times$50$\times$33~mm) 
scintillators simultaneously measured the 
time and energy information from each decay. 
The good time resolution of (1~nsec) is obtained for $E_{\gamma}>$150~keV.
Since the decay constant of the YAP is fast(30~nsec) and the slow components
of the scintillation is negligible small\cite{kataD},
the event pileup probability is  small and the time walk correction
can be performed perfectly as function of the deposited energy.

\subsection{Analysis}

The ratio $\lambda_{pick}(t)/\lambda_{3\gamma}$ is 
determined using the energy spectrum measured by 
the Ge detectors. The energy spectrum of the o-Ps decay sample, 
referred to as the {\it o-Ps spectrum}, is 
obtained by subtracting accidental contributions 
from the measured spectrum. 
The $3\gamma$-decay continuum spectrum is estimated using Monte Carlo
simulation in which the geometry and various material distributions
are reproduced in detail\cite{kataD}.
For every simulated event, three photons are generated according to 
an order-$ \alpha $-corrected energy spectrum.
Successive photoelectric, Compton, or Rayleigh scattering interactions 
of every photon are then followed through the materials 
until all photon energy is either deposited or escapes  from the detectors. 
The response function of the detectors is determined based on the measured spectrum of 
monochromatic $\gamma$-rays emitted from ${}^{152}{\rm Eu}$, ${}^{85}{\rm Sr}$, and 
${}^{137}{\rm Cs}$, with this function being used in the simulation\cite{kataD}.
These material and detector effects are taken into account precisely, and the 
{\it $3\gamma$-spectrum} is obtained. These are continuous distributions, 
and  is normalized to the observed o-Ps spectrum with 
the ratio of event numbers  within the region (480-505~keV). 

Figure 3(a) shows the o-Ps spectrum and $3\gamma$ spectrum observed in
RUN-II~(powder).
The pick-off 2$\gamma$ spectrum, which is obtained from the o-Ps spectrum after 
subtracting the $3\gamma$-spectrum, is also superimposed in the same figure. 
The $\lambda_{pick}/\lambda_{3\gamma}$ can be calculated directly from the 
ratio of the pick-off 2$\gamma$  and 3$\gamma$ spectrum, 
the ratio does not depend on the absolute values of the detection efficiencies.
The calculations of the $\lambda_{pick}/\lambda_{3\gamma}$ are
performed for the various time windows,
and its time dependence is shown in Fig.3(b).
Horizontal axis of the figure is the time between the positrons emission and 
decay, this slop show the thermalization process of the positronium.
It takes much time than the lifetime to well thermalized, as we have already shown
in the previous measurements\cite{ASAI95,JIN}.  

\begin{figure}
  \includegraphics[height=.3\textheight]{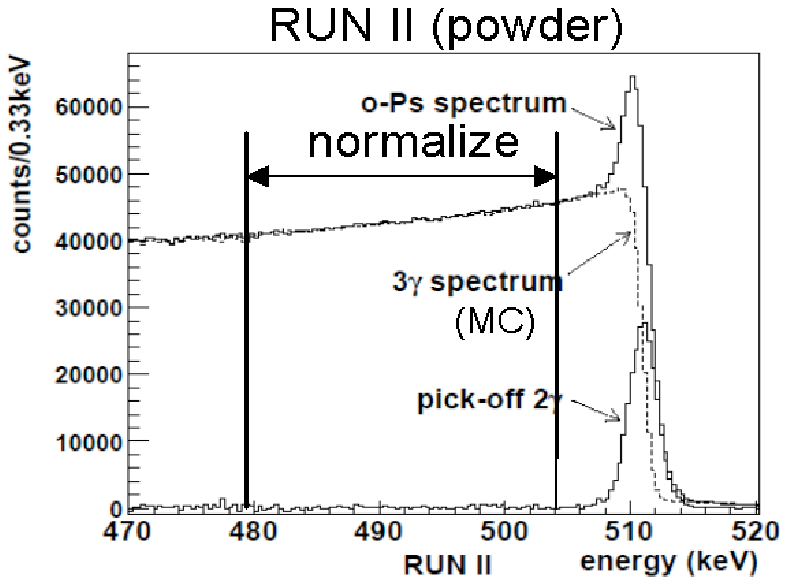}
  \includegraphics[height=.3\textheight]{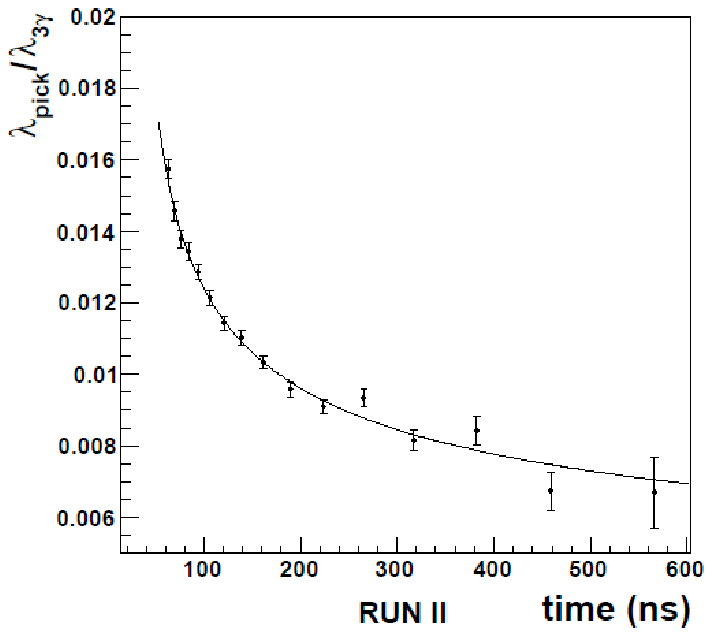}
  \caption{(a) Energy spectrum of o-Ps decay $\gamma$'s obtained by Ge detectors. 
The solid line represent data points in a time window of $60-700~nsec$, and the dotted 
line shows the $3\gamma$-decay spectrum calculated by the Monte Carlo 
simulation. Pick-off spectrum obtained after subtracting the $3\gamma$ contribution 
from the o-Ps spectrum is superimposed. 
(b) The ratio $\lambda_{pick}(t)/\lambda_{3\gamma}$ are plotted as a function of 
time for RUN-II(powder). Only the statistical errors are shown and the solid lines 
represent best fit results obtained.
}
\end{figure}

\begin{figure}
  \includegraphics[height=.3\textheight]{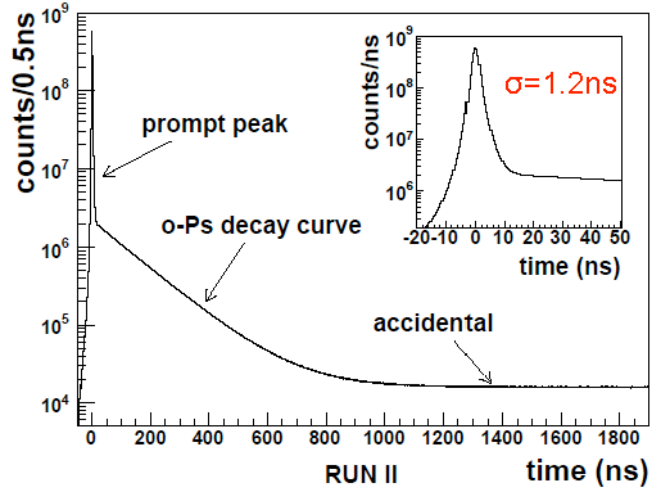}
  \caption{Timing spectrum measured by the YAP scintillators 
within an energy window above 150~keV for RUN-II.
The figure in the right corner shows
the magnified view of the prompt peak. 
}
\end{figure}

Figure~4 shows the time spectrum of the YAP scintillators 
with an energy window above 150~keV, in which a sharp peak of the prompt 
annihilation is followed by the exponential decay curve of o-Ps and then the constant 
accidental spectrum. Time resolution is 1.2 nsec is obtained and 
the o-Ps curve is widely observed over $1.0~\mu s$.
We fit resultant time spectrum using the least square method, i.e.,
\begin{equation}
N_{obs}(t)=\exp(-R_{stop}t)\left[\left(1+\frac{\epsilon_{pick}}
{\epsilon_{3\gamma}}\frac{\lambda_{pick}(t)}{\lambda_{3\gamma}}\right)N(t)
+C\right], 
\label{eq:tspec_obs}
\end{equation} 
where $\epsilon_{pick}$ and $\epsilon_{3\gamma}$ are 
the detection efficiencies for pick-off 
annihilation and $3\gamma$ decays, and $R_{stop}$ 
is an experimental random counting rate 
representing the fact that time interval measurement 
always accept the first $\gamma$ as a stop signal. 
$\lambda_{pick}/\lambda_{3\gamma}$ is about $1\%$ 
due to the low-density of the ${\rm SiO_2}$ powder or aerogel, i.e., 
the ratio of error propagation to decay rate is suppressed by a factor of 100. 

The obtained decay rates\cite{kataD} are 
$\lambda_{3\gamma}=7.03876\pm0.0009(stat.)~\mu s^{-1}$ for RUN-I and 
$7.04136\pm0.0009(stat.)~\mu s^{-1}$ for RUN-II, 
which are consistent with each other. 

Estimates of various systematic errors are summarized in Table~1 and 
the details are shown in reference\cite{kataD}.
We just mention here about errors related to simulation 
and physics source(Stark effect and 3$\gamma$ annihilation
 in the pick-off collision).

\begin{table}
\begin{tabular}{|l|r|r|}
\hline
 Source & RUN-I(ppm) & RUN-II(ppm) \\ \hline \hline
TDC non linearity & $<\pm$ 15 & $<\pm$ 15 \\ \hline
Pile up           & $< +$ 10 & $< +$ 10 \\ \hline
Pickoff estimation &    &   \\ 
 (1) 3$\gamma$ subtraction & $<\pm$ 89 & $<\pm$ 91 \\
 (2) Ge efficiency         & $<\pm$ 33 & $<\pm$ 28 \\
 (3) YAP efficiency        & $<\pm$ 64 & $<\pm$ 19 \\ \hline
Physics source &    &   \\ 
 (1) Zeeman effect         & $< -$ 5 & $< -$ 5 \\
 (2) Stark effect          & $< +$ 3 & $< +$ 4 \\
 (3) 3$\gamma$ annihilation & $< -$ 91 &$< -$ 33 \\ \hline \hline
Total     & -147 and + 115 & -104 and +98 \\ \hline 
\end{tabular}
\caption{Summary of the systematic errors}
\end{table}

The predominant contribution to total systematic error is produced 
by uncertain normalization of the 3$\gamma$ spectrum.
That is, the number of pick-off events are determined by subtracting 
the normalized 3-$\gamma$ spectrum of Monte 
Carlo simulation from the o-Ps spectrum, 
where changing the normalization factor affects the 
$\lambda_{pick}(t)/\lambda_{3\gamma}$ values and 
eventually propagates to the final result. 
Since the sharp fall-off of the 3$\gamma$-spectrum at 511~keV is solely 
produced by the good Ge energy resolution of $\sigma= 0.5$~keV, 
this subtraction only affects the lower side 
of the pick-off spectrum such that 
improper subtraction results in asymmetry of the pick-off spectrum shape. 
Comparison of the asymmetries of the pick-off peak shape 
and the prompt peak annihilation spectrum is a good parameter for estimating 
this systematic error. 
The errors are about 90~ppm for both measurements.

The Stark shift stretches the lifetime of Ps atoms, i.e., 
a probative calculation shows that the shift is 
proportional to a square of the effective electric field $E$ such that 
$\triangle\lambda_{3\gamma}/\lambda_{3\gamma}=248\cdot(E/E_0)^2$, 
where $E_0 = m_e^2e^5/\hbar^4 \approx 5.14\times10^9~V/cm$. 
Silanol functional groups on the surface of the grain 
behave as an electrical dipole moment creating an effective field around the grains. 
Average densities of Silanol are measured to be  $0.44 nm^{-2}$ and
the effective field can be analytically calculated such that the contribution to the o-Ps decay 
rate is determined to be $-5 ppm$.
These estimations were confirmed  using results from precise hyper-fine-structure (HFS)
interval measurements of ground state Ps in silica powder\cite{HFS}, 
where the interval is proportional to the size of Stark effect. 
Considering the difference in powder densities used, 
the HFS results are consistent with our estimation.

Other sources of systematic errors:
Error contribution due to the Zeeman effect is estimated
using the measured absolute magnetic field around the positronium assembly ($-5~ppm$). 
Since the 3-$\gamma$ pick-off process can only occur at a certain 
ratio, the calculated relative frequency 
$\sigma_{3\gamma}/\sigma_{2\gamma}\sim1/378$ is consistent 
with previous measurements\cite{3-pick}, 
being $-91~ppm$ for RUN-I and $-33~ppm$ for RUN-II. 

The above discussed systematic errors are regarded 
as independent contributions such that the total 
systematic error can be calculated as their quadratic sum, 
resulting in $-147~ppm$, $+115~ppm$ for RUN-I 
and $-104 ppm$, $+98 ppm$ for RUN-II.
The combined result with systematic error is 
$\lambda_{3\gamma}=7.0401\pm0.0006(stat.)^{+0.0007}_{0.0009} (sys.) ~\mu s^{-1}$,
and total error is 150~ppm.

\subsection{Results and discussion}

Figure 5 shows the summary of the measured decay rate after 1995,
three results are obtained with our method and 
one result is obtained using thin polymer, with which the produced
positronium is very slows and the the effect of the thermalized 
positronium is suppressed.
These four results are consistent with each other.
and the combined value 
is $\lambda_{3\gamma}=7.0401\pm0.0007~\mu s^{-1}$
and shown in the red arrow in the figure.
The correlations of the systematic errors 
are carefully taken into account.
The combined result is consistent with the all four results 
and accuracy is 100 ppm.
This result is consistent with the 
O($\alpha ^{2}$) correction\cite{ADKINS04},  
and differs from the only up to O($\alpha$)  by 2.6$\sigma$. 

\begin{figure}
  \includegraphics[height=.3\textheight]{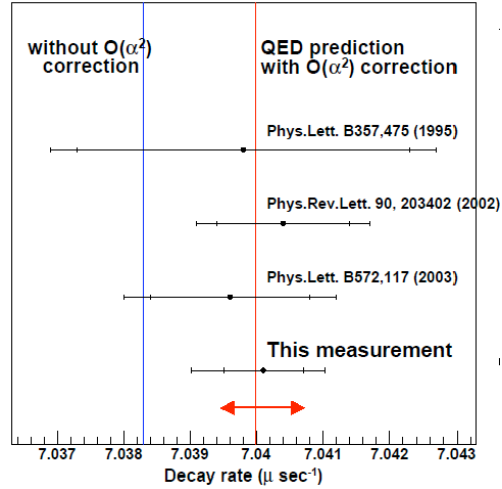}
  \caption{Decay rates measured in the last four experiments
 are listed chronically.
The last arrow shows the combined result of the last four measurements. 
The red and blue lines shows the QED prediction calculated up to O($\alpha^2$) and O($\alpha$), respectively }
\end{figure}

\section{Hyper Fine Splitting}

\subsection{Current status}

The hyper fine splitting (HFS) is the difference
of the energy level between the singlet ($1^{1}S_{0}$)
state, parapositronium(p-Ps) and the o-Ps.
The HFS of the positronium is large because of the following
two reasons:
(1) The magnetic moment is proportional to the inverse of the mass,
and large spin-spin interaction is expected for o-Ps.
(2) o-Ps has the same quantum number as photon, and o-Ps has
quantum oscillation into virtual photon,
o-Ps $\rightarrow \gamma^{*} \rightarrow$ o-Ps. 
This frequency of 87GHz contributes only to the o-Ps, 
and makes HFS larger. 

The precise measurement of the HFS gives the direct information about
the higher order calculation of the QED, especially of the bound state QED.
If an unknown light particle (like as axion or millichaged particle) exits,
it contributes to the energy level, and makes discrepancy from the QED prediction.
Since the quantum oscillation has good sensitivity to the hypothesis 
particle, whose coupling is super-weak, the precise measurement 
of HFS is good tool to search for the new physics beyond
the Standard Model indirectly.

The precise measurements have been performed in 70's and 80's.
These results are consistent with each other and the final 
precision is 3.6 ppm~\cite{HFSnew}.
No higher order correction($\geq O(\alpha^{2})$)
has been performed at that time and these results ``were'' 
consistent with the 1st order calculation of the QED.
New method to calculate the higher order correction on the bound state 
is established in 2000, and the 
2nd and 3rd corrections have been performed\cite{HFSth}.
The QED prediction is 203.3917(6) GHz and it differs from the measured value of
203.38910(74) GHz. 
The discrepancy of 3.5$\sigma$ is observed.
It is still marginal for statistical fluctuation,
but there would be possibility of the new physics or
the common systematic errors in the previous measurements.
(There is also possibility that NRQED approximation is wrong.)

\subsection{Old experiments and the systematic errors}
 
In the previous all measurements, the HFS transition was not directly measured,
since 203GHz was too high frequency to be handled.
The static magnetic field makes Zeeman mixing between $m_z$=0 state of o-Ps 
and p-Ps, the resultant energy level of $m_z$=0 state
becomes higher than  $m_z=\pm$1. 
This energy shift is proportional to the HFS energy level. 
Figure~6 shows the energy levels as function of the static magnetic field.
The $m_z=\pm$1 state of o-Ps do not couple to the static magnetic field,
and do not change the energy level.

\begin{figure}
  \includegraphics[height=.3\textheight]{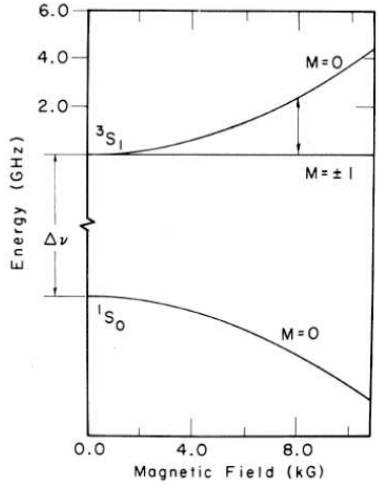}
  \caption{The energy level of Ps as function of the static magnetic field. 
$\Delta \nu$ represents the HFS, and o-Ps state ($^{3}S_{1}$) is higher 
than p-Ps ($^{1}S_{0}$). 
Arrow shows the Zeeman shift between $M_z=0$ and $M_z=\pm1$ at B=8kGauss.}
\end{figure}

Positronium was produced with the $\beta^+$ source and gas($N_2$, Argon etc)
in the RF cavity, in which the high power 2.3GHz microwave was stored.
Static magnetic field (about 8kGauss shown with arrow in the figure) 
was also applied and scanned near the resonance of the Zeeman shift.
When the Zeeman energy shift is just on the RF frequency, the transition from
$m_z=\pm$1 to $m_z$=0 increases, and the $m_z$=0 state decays into
2$\gamma$ immediately through p-Ps state.
The 2$\gamma$ decay, which were tagged with back-to-back topology, 
increases on the resonance.
HFS can be determined as center value of the resonance peak 
by scanning the magnetic field.
Since Ps is produced in the gas, the Ps collides with the gas molecule and
the electric field of the gas molecule makes the shift of the energy 
state, which is called as the Stark effect, 
and about 10ppm shift was observed for 1 atm gas\cite{HFSnew}. 
The HFS in gas were measured by changing the gas pressures. 
The measured values were extrapolate to zero pressure,
and the HFS in the vacuum was obtained. 
This extrapolation method was the exactly 
same as in the decay rate measurements\cite{GAS87,GAS89,CAV90}. 

There are two possibilities of systematic errors in this method.
(1) First is the effect of the non-thermalized o-Ps, which is
the same as in the decay rate measurement.
Extrapolation procedure is assumed that Ps is well thermalized and 
the mean velocity of the Ps is the same for the various pressured gases.
We have already shown in the decay rate measurements
that this well-thermalized assumption is not satisfied, 
and that this method makes big systematic errors. 
The non-thermal Ps would affect also on the HFS measurement.
(2) The Ps is widely spread in the RF cavity (size 17cm in diameter) 
and non-uniformity of the static magnetic field was the main 
source of the uncertainties of the results.
Size of the used magnet were limited and 
the uniformity of the filed was about 10ppm level, which has been 
corrected. 

\subsection{Our new methods}

We propose the following two new methods in order to 
solve these systematic uncertainties in the previous measurements.

\subsubsection{Conventional Zeeman method using the magnet field}

$^{22}Na$ positron source(1MBq) is installed in the 
thin plastic scintillator(100$\mu$m),
and the timing information of the positron emission
is tagged (positron emits at t=0).
There are two benefits to have time information:
(1) Many part of positron is just annihilated into 2$\gamma$ at t=0,
and this 2$\gamma$ annihilation events make S/N seriously worse.
The 2$\gamma$ annihilation background events can be removed 
dramatically with the requirement of t$>$10nsec.
(2) Thermalization process can be determined with the energy spectrum 
measured by the germanium semiconducting detectors, 
as the same as in the measurement of the decay rate.
Stark effect (material effect) can be 
measured changing gas pressure and 
the non-thermalization effect are measured directly.
    
Magnetic filed is provided with the superconducting
magnet developed for the medical NMR.
Figure~7 shows the photograph of the magnet, which
has the large bore size (80cm) and strong (about 1T) magnetic field.
The magnetic filed is uniform well, 
we can reduce the uncertainties due to non-uniformity of the magnetic filed.
0.9T of the magnetic filed will be applied with 2.8GHz RF (power 500W)
microwave.
Cavity with TM110 mode resonance is used to store the RF power.

\begin{figure}
  \includegraphics[height=.3\textheight]{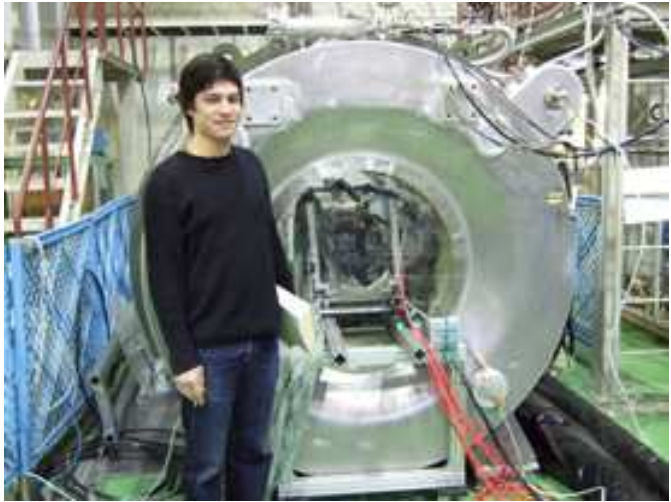}
  \caption{Superconducting magnet: Bore size is 80 cm and
   the uniform magnetic field can be applied up to 1T.}
\end{figure}

Figure 8(a) shows the gamma ray detectors and the RF cavity.
Six LaBr$_3$ scintillators and two germanium detectors
are used to detect gamma ray from the Ps decay.
LaBr$_3$ scintillator has the fast signal shape as shown in Fig.8(b).
Fast raising time, faster than 10 nsec, provides the good time resolution of
about 300psec (FWHM at 511KeV),
and the fast decay time (26nsec) grantees 
the pileup effect can be reduced.
Energy can be measured with the high resolution (FWHM=3\% at 511keV) 
as shown in Fig.8(b).
The energy resolution is better than NaI by factor 2,
then 2$\gamma$ processes can be tagged 
with the single scintillator using the energy information
(Energy tagging).
Since the tagging efficiency of the energy tagging is significantly higher than
the `geometrical tagging', in which 2$\gamma$ decay
is tagged with two scintillators located back-to-back.
The expected numbers are summarized in table 2.
First measurement with this new method will be performed in this Autumn, 
and a few ppm accuracy will be obtained for a half year.

\begin{figure}
  \includegraphics[height=.3\textheight]{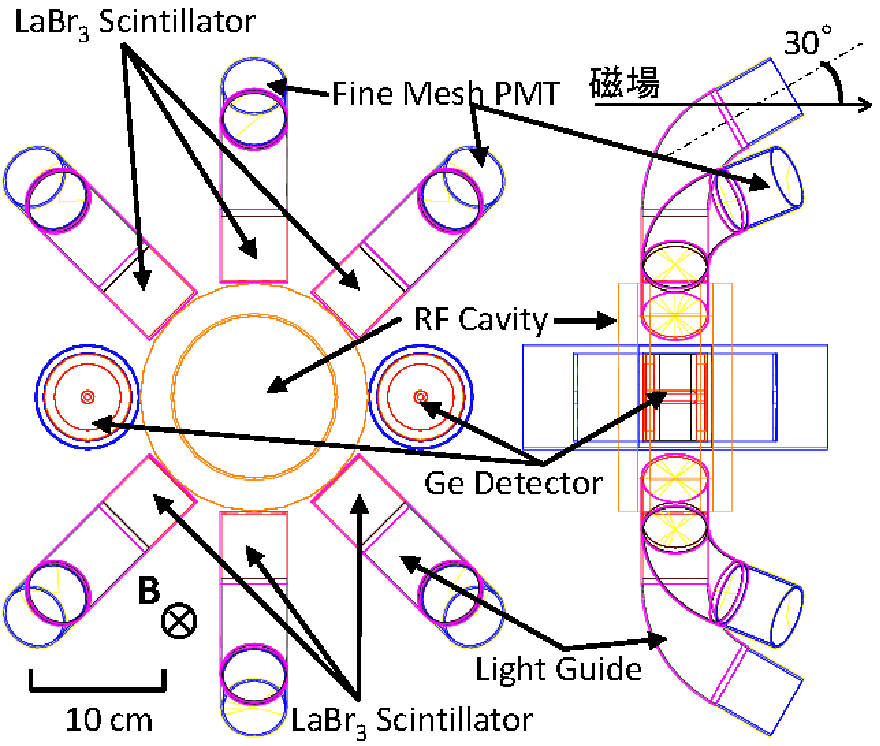}
  \includegraphics[height=.3\textheight]{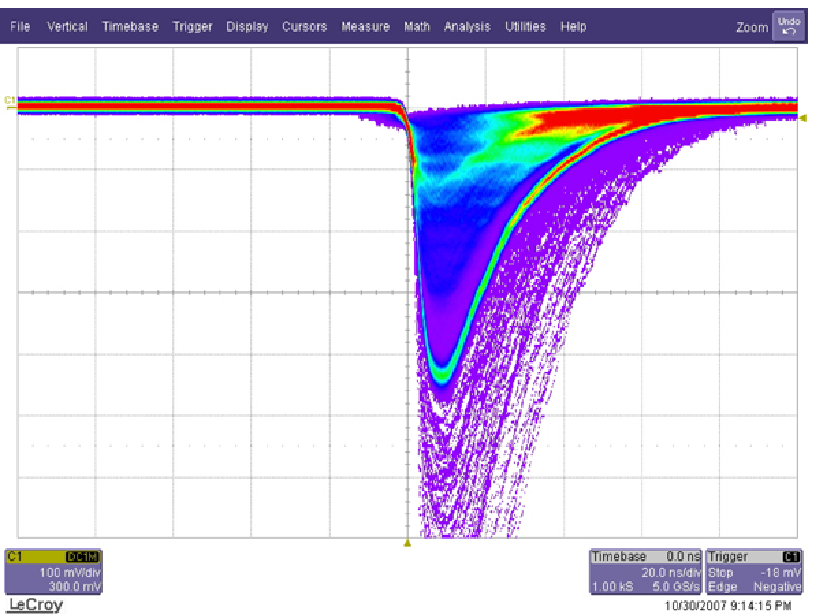}
  \caption{(a) The experimental setup of the gamma ray detectors:
  6 LaBr$_{3}$ scintillators and 2 Ge detectors are located around 2.8GHz RF cavity.
  Since these are located in the magnetic filed, the fine-mesh PMTs are
  used and the angel between the magnetic field 
  and the PMT axis is set at 30$^{o}$, at which the magnetic field effect
  on the PMT gain is about 10. (b) The observed signal shape 
   of the LaBr$_{3}$ for 
  667KeV gamma ray emitted from $^{137}$Cs.} 
\end{figure}

\begin{table}
\begin{tabular}{|l|r|r|}
\hline
                & geometrical tagging & Energy tagging \\ \hline 
2$\gamma$ decay & 4.3$\times 10^4$ & 3.2$\times 10^6$  \\ \hline
3$\gamma$ decay &  7$\times 10^3$  & 2$\times 10^6$   \\ \hline
\end{tabular}
\caption{The expected numbers of events for 2$\gamma$ and 3$\gamma$ decay with
the both tagging method. Numbers will be detected on the Zeeman resonance 
for one day run}
\end{table}

\subsubsection{Direct transition using sub THz RF}

(Sub) THz RF is the unexplored filed, since (sub) THz RF has the 
characters between the optics and the radio wave.
Developing the high quality/high power source for the  (sub) THz RF 
is useful and interesting for both science and technology.
We can observe the direct transition of the HFS without magnet field,
if we have good source of 203GHz RF.
It is free from the systematic error due to magnetic filed, and
it open the new paradigm of the atomic/particle physics 
using (sub) THz. 

Figure9(a) shows photograph of the gyrotron RF generator\cite{THZ}, 
in which high power(>100W) sub THz
RF can be produced continuously.
The accelerated electron emits RF in the strong magnetic filed, 
and the one/specific frequency is enhanced with the resonance cavity. 
Arbitrary RF frequency can be picked up with the tuning resonance cavity.
Tunable frequency is also big challenge of THz RF source. 
Improvement stability of the power is in progress and 
the final design of stability is about 10~ppm.
This is big challenge on the THz technology
and we can expect various spin off for not only basic science
but also technology.

203 $\pm$ 1GHz (tunable frequency) CW RF source 
will be ready in this year.
The gamma ray detectors mentioned above are used, 
and the positron timing 
is also measured in order to reduce annihilation 2$\gamma$ and measure 
the thermalization process.
The Fabry-Perot cavity is used for RF storage, 
since 203GHz has optical characters.
Figure 9(b) shows the expected resonance curve with 100W RF power 
and Q=6000 cavity.
The width(FWHM 0.63GHz) is determined with the decay rate of p-Ps, and 
the center value can be determined with accuracy of about 100~ppm.

\begin{figure}
  \includegraphics[height=.3\textheight]{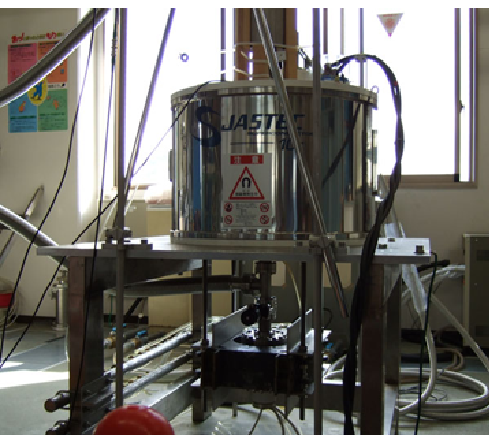}
  \includegraphics[height=.3\textheight]{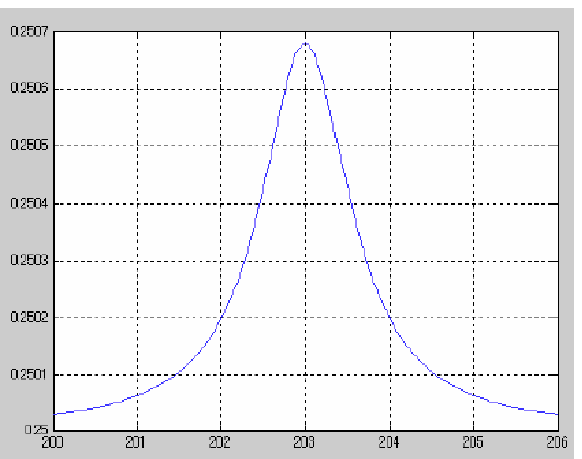}
  \caption{(a) Photograph of Gyrotron FU CW-4: High power (500W) and 
  high frequency (400GHz) sub THz wave can be produced.
  (b) The expected resonance curve with Q=6000 cavity and 100W RF source. 
}
\end{figure}

\section{Conclusion}

The decay rate of o-Ps was measured using a 
direct $2\gamma$ correction method in which the thermalization 
effect of o-Ps is accounted for and integrated into the time spectrum 
fitting procedure. 
The new result is 
$\lambda_{3\gamma}=7.0401\pm
0.0006(stat.)^{+0.0007}_{-0.0009} (sys.) ~\mu s^{-1}$,
and total error is 150~ppm\cite{kataD}.
It is most accurate result and agrees well 
with the last three results\cite{ASAI95,JIN,MPOL}. 
The combined result of these four measurements
is 
7.0401$\pm$0.0007$\mu s^{-1}$,
which is consistent well with th NRQED prediction 
corrected up to $O(\alpha^2)$ term\cite{ADKINS-4},
and differ from the result up to $O(\alpha)$ by 
2.6$\sigma$\@.
It is the first result to validate the $O(\alpha^{2})$
correction.

The Hyper Fine Splitting between o-Ps and p-Ps 
is sensitive to the higher order corrections 
of the QED calculation, 
and the discrepancy of 3.5$\sigma$ is found between 
the measured values and the 3rd order QED prediction.
We point out that there are two possibilities 
of the common systematic errors 
in the previous all measurements.
(1) Non-thermalized Ps makes error in the extrapolation method
as the same as in the decay rate measurements.
This was serious source of the orthopositronium lifetime puzzle.
(2) Non-uniformity of the magnetic field is also serious source
of the systematic error.

We propose new methods to measure HFS precisely without 
the these uncertainties.
Thermalization processes are measured with the $2\gamma / 3\gamma$ 
ratio, and also 2$\gamma$ annihilation background events can be reduced 
by requirement of t > 10~nsec.
LaBr$_3$ scintillators are used for tagging 2$\gamma$ decay,
since its energy resolution is better than the conventional 
scintillator {\it i.e.} NaI and CsI, 
we can tag 2$\gamma$ decay only with the energy information,
not with back-to-back topology. It gains by factor 70 for 
the tagging efficiency.
Large bore superconducting magnet is prepared for 
new measurement, it provides the uniform/strong magnetic filed.

203GHz high power and stable RF source are developed.
It makes possible to transit p-Ps into o-Ps directly.
We can measure HFS directly instead of Zeeman shift.
It will be the first application of sub THz RF source   
to the basic science.

\bibliographystyle{aipproc}   


\begin{thebibliography}{99}

\bibitem{GAS87}
C. I. Westbrook, D. W. Gidley, R. S. Conti, and A. Rich, 
\emph{Phys. Rev. Lett} \textbf{58} 1328 (1987).

\bibitem{GAS89}
C. I. Westbrook, D. W. Gidley, R. S. Conti, and A. Rich, 
\emph{Phys. Rev.} \textbf{A40} 5489 (1989).

\bibitem{CAV90}
J. S. Nico, D. W. Gidley, A. Rich, and P. W. Zitzewitz, 
\emph{Phys. Rev. Lett.} \textbf{65} 1344 (1990).

\bibitem{ADKINS-4}
G. S. Adkins, R. N. Fell, and J. Sapirstein, 
\emph{Phys. Rev. Lett.} \textbf{84} 5086 (2000) and
\emph{Ann. Phys.} \textbf{295} 136 (2002).

\bibitem{PEKIN}
S. Asai , T. Hyodo, Y. Nagashima, T.B. Chang and S. Orito, 
\emph{Materials Science Forum,} \textbf{619} 175 (1995).

\bibitem{ASAI95}
S.Asai \emph{New measurement of orthopositronium lifetime}, 
\textbf{Ph. D. thesis}, University of Tokyo (1994);
S. Asai, S. Orito, and N. Shinohara, \emph{Phys. Lett.} 
\textbf{B357} 475 (1995).

\bibitem{JIN}
O.~Jinnouchi, \emph{Study of bound state QED: precision measurement of the
  orthopositronium decay rate}, \textbf{Ph. D. thesis}, 
University of Tokyo (2001);
O. Jinnouchi, S. Asai, and T. Kobayashi, \emph{Phys. Lett.} 
\textbf{B572} 117 (2003). 

\bibitem{MPOL}
R.S.~Vallery, P.W.~Zitzewitz and D.W.~Gidley,
\emph{Phys. Rev. Lett} \textbf{90} 203402 (2002).


\bibitem{kataD}
Y.~Kataoka \emph{Test of bound State QED: Higher order correction:
Precision measurement of orthopositronium decay rate}, 
\textbf{Ph. D. thesis}, University of Tokyo (2007),
\url{http://tabletop.icepp.s.u-tokyo.ac.jp/oPs-life/main.final.pdf}

\bibitem{HFS}
M.H.~Yam, P.O.~Egan, W.E.~Frieze, and V.M.~Hughes, 
\emph{Phys.~Rev.} \textbf{A18} 350 (1978). 

\bibitem{3-pick}
J.~A.~Rich, \emph{Phys.~Rev} \textbf{61} 140 (1951).

\bibitem{HFSnew}
M.W.~Ritter, P.O.~Egan, V.W.Highes and K.A.Woodle,
\emph{Phys. Rev.} \textbf{A30} 1331 (1984).
A.P.~Mills. Jr, \emph{Phys. Rev.} \textbf{A27} 262 (1983). 

\bibitem{HFSth}
G.S.~Adkins, R.N.~Fell, and P.Mitrikov.
\emph{Phys. Rev. Lett.} \textbf{79} 3387 (1997).
B.~kniehl and A.A.Penin, 
\emph{Phys. Rev. Lett.} \textbf{85} 5094 (2000).

\bibitem{THZ}
T. Idehara et al. \emph{Int. J. of Infrared and Millimeter Waves}
\textbf{28} 433 (2007).

\end{thebibliography}


\end{document}